\documentclass[aps,prl,showpacs,twocolumn,amsmath,amssymb,superscriptaddress,footinbib,epsfig]{revtex4}

\usepackage[english]{babel}

\usepackage{latexsym}
\usepackage{graphicx}
\usepackage{epsfig}
\usepackage{amsfonts}
\usepackage{amssymb}
\usepackage{amsmath}
\usepackage{bbm}

\begin{document}

\title{A Kochen-Specker inequality from a SIC}

\author{Ingemar Bengtsson}
\author{Kate Blanchfield}
\affiliation{Stockholms Universitet, Fysikum, S-10691 Stockholm,
Sweden}
\author{Ad\'{a}n Cabello}
\affiliation{Departamento de F\'{\i}sica Aplicada II, Universidad de Sevilla, E-41012
Sevilla, Spain}
\affiliation{Stockholms Universitet, Fysikum, S-10691 Stockholm, Sweden}

\date{\today}

\begin{abstract}
Yu and Oh [1] have given a state-independent proof of the Kochen-Specker theorem 
in three dimensions using only 13 rays. The proof consists of showing that a non-contextual 
hidden variable theory necessarily leads to an inequality that is violated by quantum 
mechanics. We give a similar proof making use of 21 rays that constitute a SIC 
and four Mutually Unbiased Bases.

\end{abstract}

\pacs{03.65.Ta, 03.65.Ud}
\maketitle

{\it Introduction.---}The Kochen-Specker theorem states that a certain kind of hidden variable 
theory cannot be consistent with quantum mechanics. The idea is to assign truth values (1 for 
true, 0 for false) to a finite set of measurements represented by projectors onto rays in 
Hilbert space. These assignments must obey the Kochen-Specker rules, namely no two 
orthogonal projectors can both be 
true, and one member of each complete orthonormal basis must be true. Since two orthogonal 
projectors commute they represent compatible measurements. Note that the assignment made for 
a particular projector is independent of which particular set of mutually compatible 
measurements it belongs to---even though it may belong to several such contexts. Such a
hidden variable theory is said to be non-contextual. 

The usual proof proceeds by finding a finite set of projectors with a pattern 
of orthogonalities such that there does not exist a truth value assignment consistent with the rules 
\cite{KS, Peres}. In other words, a 
non-contextual hidden variable theory reproducing the quantum mechanical results is shown 
to be logically impossible. The smallest number of projectors needed for such a proof 
seems to be 18 (in four dimensions \cite{Cabello}), or 31 (in three dimensions \cite{Conway}).   

Klyachko \textit{et al.} \cite{Can} noticed that, using only five projectors in three dimensions, 
the Kochen-Specker rules lead to an inequality for observed frequencies that can be violated 
in quantum mechanics, if a particular quantum state is chosen. It was further observed that 
one can find a set of projectors so that, without employing the 
Kochen-Specker rules, they lead to an inequality 
violated by all quantum states (including the totally mixed state) \cite{Adan}. In fact any version of the usual proof leads to such an 
inequality \cite{Piotr}. We call the first type of inequality a Kochen-Specker inequality 
and the second type, where the truth value assignments are constrained only by the assumption of non-contextuality, a non-contextual inequality. Yu and Oh \cite{Oh} 
found a state-independent Kochen-Specker inequality from 4 projectors chosen from 
a larger set of 13, and a state-independent non-contextual inequality from the same set.

It is difficult to prove experimentally that something is logically impossible. On the 
other hand the reformulation of the Kochen-Specker theorem in terms of inequalities has 
led to a number of recent experimental tests \cite{nagon, nagon2, Elias, Moussa, Anton}. 
Using inequalities also has the incidental advantage that the Kochen-Specker theorem 
can be proved over the rational numbers \cite{Larsson}.  

Our purpose is to give a state-independent proof along the same lines as Yu and Oh, 
but starting from a configuration of rays in three dimensions that is of independent 
interest: a symmetric informationally-complete POVM (SIC) and a complete set of mutually unbiased bases 
(MUB). The resulting configuration of 21 rays is highly symmetric, and we believe 
that it has some advantages.

{\it Our 21 vectors.---} Let $q = e^{2\pi i/3}$, a third root of unity. 
In unnormalised form the first nine vectors are 

\begin{equation} \begin{array}{cccr} (0,1,-1) \ & (0,1,-q) \ & (0,1,-q^2) \ \\ \\ \\ 
(-1,0,1) \ & (-q,0,1) \ & (-q^2,0,1) \ \\ \\ 
\\ (1,-1,0) \ & (1,-q,0) \ & (1,-q^2,0 ) \ & \ . \end{array} 
\label{SIC} \end{equation}

\noindent These vectors form a POVM, a set of vectors such that 
if one sums all the projectors $|\psi_i\rangle \langle \psi_i|$ one obtains an operator 
proportional to the unit matrix. In fact they form a SIC \cite{Zauner, Renes}. 
For our purposes, a SIC in dimension $N$ is simply a collection of $N^2$ unit vectors 
such that 

\begin{equation} \sum_{i=1}^{N^2}|\psi_i\rangle \langle \psi_i| = N{\mathbbm 1} \ ,
\label{completeness} \end{equation}
\begin{equation} |\langle \psi_i|\psi_j\rangle |^2 = \frac{1}{N+1} \hspace{5mm} 
\mbox{if} \hspace{5mm} i \neq j \ . \end{equation}

\noindent A SIC is a very special kind of POVM. There is much more to say about the 
two notions 
we just introduced---they are used to describe measurements of a more general kind 
than the usual von Neumann measurements---but in this letter none of this is relevant. 
The measurements we are thinking of are ordinary projective measurements. 

We use twelve further vectors that can be collected into four bases. Again in 
unnormalized form they are the columns of the matrices 

\begin{eqnarray} \Delta^{(0)} = \left[ \begin{array}{ccc} 1 & 0 & 0 \\ 0 & 1 & 0 \\ 
0 & 0 & 1 \end{array} \right] \ , \hspace{6mm} \Delta^{(\infty )} = \left[ \begin{array}{ccc} 
1 & 1 & 1 \\ 1 & q & q^2 \\ 
1 & q^2 & q \end{array} \right] \ , \nonumber \\ \\ 
\Delta^{(1)} = \left[ \begin{array}{ccc} 1 & q^2 & q^2 \\ q^2 & 1 & q^2 \\ 
q^2 & q^2 & 1 \end{array} \right] \ , \hspace{6mm} \Delta^{(2)} = \left[ \begin{array}{ccc} 
1 & q & q \\ q & 1 & q \\ 
q & q & 1 \end{array} \right] \ . \nonumber \end{eqnarray} 

\noindent These four bases have a particular relation to each other, and form 
a complete set of MUB \cite{Ivanovic, Wootters}. Again this is 
irrelevant here.

What is relevant is the special relationship that the SIC and the MUB enjoy in 
three dimensions. As was noted some time ago \cite{Hesse}, each of the twelve 
MUB vectors is orthogonal to three SIC vectors in such a way that each basis divides 
the SIC vectors into three distinct sets. Moreover each SIC vector is
orthogonal to four MUB vectors, one from each basis. This pattern of orthogonalities 
is sometimes referred to as the Hesse configuration; we will comment on this later. 

{\it A state-independent Kochen-Specker inequality.---}Our first inequality will concern the SIC vectors 
only, but we constrain their truth values by first assigning truth values to the basis 
vectors. The Kochen-Specker rules force us to distribute 
one ``1'' and two ``0s'' among the three vectors in each of the four bases. Because 
each such vector is orthogonal to three SIC vectors this will force us to assign 
``0'' to some of the latter. 

Let us see how this works. We think of the SIC vectors 
as nine points in a square array arranged as in (\ref{SIC}). In the figures 
the lines represent basis vectors. 
The three SIC vectors orthogonal to a given basis vector lie on the corresponding line.  
A solid line represents a basis vector assigned ``1'', and the 
orthogonalities then force the SIC vectors on that line 
to be assigned ``0'' (as shown by filled dots). In Fig. 1 we do this first for 
$\Delta^{(0)}$ and then for $\Delta^{(\infty )}$. SIC vectors whose assignments are 
still undetermined are shown as empty dots. 

\begin{figure}[ht]
\centerline{\includegraphics[width=0.60\linewidth]{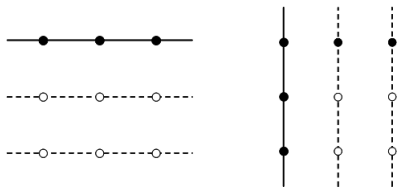}}
\caption{Truth value assignments from $\Delta^{(0)}$ and $\Delta^{(\infty )}$.}
\end{figure} 

For readers familiar with the Kochen-Specker literature, we emphasise that these 
are not orthogonality graphs \cite{KS}. 

We had a choice when we assigned truth values to the MUB vectors, but by the 
symmetry of the problem the choices made for the first two bases did not matter. 
When we proceed to $\Delta^{(1)}$ and $\Delta^{(2)}$ the choices do matter, 
but there are only nine possible choices and one quickly goes through them 
all. One possibility is shown in Fig. 2. Some of the ``lines'' now look curved; for 
readers familiar with such things we remark that they are in fact lines in a 
finite affine plane \cite{Wootters}.  

\begin{figure}[ht]
\centerline{\includegraphics[width=0.60\linewidth]{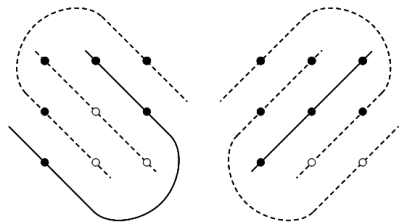}}
\caption{Truth value assignments from $\Delta^{(1)}$ and $\Delta^{(2)}$.}
\end{figure} 

In this case seven SIC vectors are assigned the value ``0'', and two SIC vectors 
are left undetermined. This will happen for eight out of the nine choices. In 
the remaining case all SIC vectors are assigned ``0''.

Using a single projective measurement we can determine the truth value of at 
most one SIC vector. Let $T_i$ be the truth value of the SIC vector 
$|\psi_i\rangle$. When we repeat the experiment 
many times we divide the ensemble into nine subensembles. Assuming that this 
does not affect the averages, we conclude that their sum 
cannot exceed 2. Thus 

\begin{equation} \sum_{i=1}^9 \langle T_i\rangle \leq 2 \ . \label{KS1} \end{equation}

\noindent This is the prediction from any non-contextual hidden variable theory. 
But the quantum mechanical average, if the state of the system is $\rho$, 
is 

\begin{equation} \sum_{i=1}^9\mbox{tr} \rho |\psi_i\rangle \langle \psi_i| 
= \mbox{tr} \rho\sum_{i=1}^9|\psi_i\rangle \langle \psi_i|
= 3 \ , \end{equation}

\noindent by eq. (\ref{completeness}). Should experiments confirm this, the hidden 
variable theories have been falsified. 

{\it A state-independent non-contextual inequality.---}It is possible to write down 
an inequality that does not assume the Kochen-Specker rules, and which is violated by 
quantum mechanics. The only assumption made in deriving such a non-contextual inequality
is that the hidden variable theories assign values to 21 dichotomic observables in a 
non-contextual manner.  

We choose to represent the 21 quantum mechanical observables by the 21 operators 

\begin{equation} A_i = {\mathbbm 1} - 2|\psi_i\rangle \langle \psi_i| \ , \end{equation}

\noindent with the spectrum $(1,1,-1)$. We also introduce the dichotomic hidden 
variables $a_i$, taking the values $\pm 1$. 

Let us introduce the symbol 
$\Gamma_{ij}$, $1 \leq i,j \leq 21$, which by definition is equal to 1 if 
the measurements labelled $i$ and $j$ are compatible and distinct (that is if 
$A_i$ and $A_j$ commute), and equal to 0 otherwise. We are interested in the function 

\begin{equation} C = \sum_{i=1}^{21}a_i -  
\frac{1}{5}\sum_{i=1}^{21}\sum_{j=1}^{21}\Gamma_{ij} a_ia_j \ . 
\end{equation} 

\noindent There are $2^{21}$ different assignments one can make for the $a_i$'s, 
but since our configuration is very symmetric an exhaustive search is not 
difficult. The conclusion is that $C$ takes values no larger than $\frac{63}{5}$. Hence when 
we take averages, the hidden variable theory predicts that

\begin{equation} \sum_{i} \left\langle A_i \right\rangle -  
\frac{1}{5}\sum_{i,j}\Gamma_{ij} \left\langle A_iA_j \right\rangle \leq \frac{63}{5} \ . 
\end{equation}

In quantum mechanics this quantity is evaluated as the expectation value of the operator 

\begin{equation} Q = \sum_{i}A_i - \frac{1}{5}\sum_{i,j}\Gamma_{ij} A_iA_j 
= \frac{67}{5} {\mathbbm 1} \ . 
\end{equation} 

\noindent A minor calculation is needed to show this. Hence the quantum expectation 
value is again independent of the state, and given by 

\begin{equation} \sum_i\mbox{tr}\rho A_i -\frac{1}{5}\sum_{i,j}\Gamma_{ij} 
\mbox{tr}\rho A_iA_j = 
\mbox{tr} \rho Q = \frac{67}{5} \ . \end{equation}

\noindent This is an obvious violation of the prediction from the 
non-contextual hidden variable theories, which is that this average should be 
less than or equal to $\frac{63}{5}$.

{\it Comments.---}We end with four comments.

{\it i}) The inequality we have presented is not unique. We could have considered the 
quantity 

\begin{equation} C = \sum_{i}a_i -  
k\sum_{i,j}\Gamma_{ij} a_ia_j \ , \hspace{6mm} \frac{1}{8} < k < \frac{1}{4} \ , 
\end{equation} 

\noindent and we would still have obtained a violation. We can also weight 
the individual terms differently. Such considerations are important if one wants an experimentally robust inequality to test.

{\it ii}) The 13 vectors appearing in the proof by Yu and Oh \cite{Oh} are real. They include the computational 
basis, two triplets of vectors that can be extended to complete SICs in complex Hilbert space, 
and a set of four vectors that are unbiased with respect to the computational basis. 
The latter four form a POVM, and play the role that the SIC vectors play in our construction. 

{\it iii}) The Hesse configuration is normally not thought of as a pattern of orthogonalities, but as 
a collection of nine vectors and twelve two-dimensional subspaces (normal to the MUB vectors), 
such that each subspace contains three vectors and each vector belongs to four subspaces. 
It is interesting to observe that a somewhat similar configuration of vectors and subspaces, 
known as the Reyes configuration, is relevant \cite{Aravind} to Peres' 24-ray proof \cite{Peres} 
of the Kochen-Specker theorem in four dimensions.

{\it iv}) The close connection between SICs and mutually unbiased bases that we have used 
appears to be special to three dimensions \cite{Appleby, Scott}. At the same time we 
note that there is a foundational aspect to SICs \cite{Fuchs}. We do not know 
if our observation can be employed in that context.

\begin{acknowledgments} We thank \AA sa 
Ericsson for discussions. IB is supported by the Swedish Research Council under contract 
VR 621-2010-4060. AC is supported by the MICINN Project No: FIS2008-05596 and the Wenner-Gren Foundation.

\end{acknowledgments}

\end{document}